\documentclass[12pt]{iopart}

\usepackage{graphicx}
\usepackage{setstack} 
\usepackage{iopams}
\usepackage{color}
\usepackage{soul}
\usepackage{mathrsfs}

\usepackage{amssymb}

\begin{document}

\title[Search reliability and search efficiency of combined L\'evy-Brownian
motion]{Search reliability and search efficiency of combined L\'evy-Brownian
motion: long relocations mingled with thorough local exploration}

\author{Vladimir V. Palyulin$^{\dagger}$, Aleksei V. Chechkin$^{\ddagger,\P,\S}$,
Rainer Klages$^{\flat,\P}$ and Ralf Metzler$^{\sharp,\pounds}$}

\address{$\dagger$ Physics Department, Technical University of Munich,
D-85747 Garching, Germany\\
$\ddagger$ Akhiezer Institute for Theoretical Physics NSC KIPT,
Kharkov, 61108, Ukraine\\
$\P$ Max-Planck-Institut f\"ur Physik komplexer Systeme, D-01187 Dresden, Germany\\
$\S$ Department of Physics \& Astronomy, University of Padova,  35122 Padova,
Italy\\
$\flat$ Queen Mary University of London, School of Mathematical Sciences,
Mile End Road, London E1 4NS, UK\\
$\sharp$ Institute for Physics \& Astronomy, University of Potsdam,
D-14476 Potsdam-Golm\\
$\pounds$ Department of Physics, Tampere University of Technology, 33101 Tampere,
Finland}
\ead{rmetzler@uni-potsdam.de}

\begin{abstract}
A combined dynamics consisting of Brownian motion and L\'evy flights is exhibited
by a variety of biological systems performing search processes. Assessing the
search reliability of ever locating the target and the search efficiency of doing
so economically of such dynamics thus poses an important problem. Here we model
this dynamics by a one-dimensional fractional Fokker-Planck equation combining
unbiased Brownian motion and L\'evy flights. By solving this equation both
analytically and numerically we show that the superposition of recurrent Brownian
motion and L\'evy flights with stable exponent $\alpha<1$, by itself implying zero
probability of hitting a point on a line, lead to transient motion with finite
probability of hitting any point on the line. We present results for the exact
dependence of the values of both the search reliability and the search efficiency
on the distance between the starting and target positions as well as the choice
of the scaling exponent $\alpha$ of the L\'evy flight component.
\end{abstract}

\maketitle

\section{Introduction}

One of the most fundamental questions about any type of motion is whether a
moving particle starting from a point A is able to reach some pre-selected
point B \cite{hughes}. After the concept of random walks was brought to wide
attention by Karl Pearson in 1905 \cite{pearson1905}, for simple random walks
on a lattice George P\'olya gave an answer to this question by proving that in
one and two dimensions random walks are recurrent, which implies that a walker will
visit any lattice site eventually. In three and higher dimensions this motion
becomes transient \cite{polya1921}. For random walks where the jump lengths
 $\ell$ are drawn from L\'evy $\alpha$-stable distributions, $p(\ell)\sim
|\ell|^{-1-\alpha}$ with $0<\alpha<2$ \cite{Levybook,Bouchaud}, the motion
is recurrent for $\alpha\ge1$, otherwise it is transient \cite{spitzer}.
However, even if the motion is transient there might be a non-zero
probability of visiting a certain point. For instance the Brownian motion
is transient in three dimensions, however, there exists a finite probability of
return.\footnote{On a simple cubic lattice in three dimensions the returning
probability is $\approx0.34$ \cite{hughes}.}
The probability of hitting a given point in space can be directly calculated
from the density $\wp_{\mathrm{fa}}(t)$ of the {\em first arrival time},
or hitting time $t$ to a specific point, which conveniently characterises
the hitting process. Integrating $\wp_{\mathrm{fa}}(t)$ over time produces
the cumulative probability $P$ of reaching this point, called the {\em search
reliability} \cite{PNAS14,LevyLong}.  

\begin{figure}\center
\includegraphics[height=4cm]{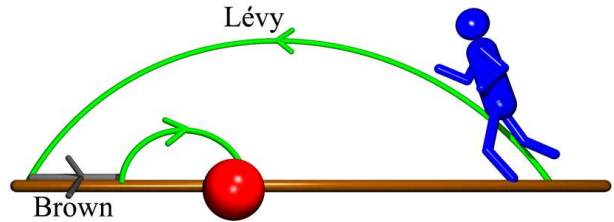}
\caption{Sketch of intermittent search in one dimension: a searcher (blue) proceeds
by a combination of L\'evy jumps and Brownian steps until it finds the target (red).
Physically, L{\'e}vy jumps decorrelate the motion, leading the searcher to sites
not previously visited. Brownian motion, instead, provides a thorough local
search at the price of oversampling, see text.}
\label{sketch1}
\end{figure}

The search reliability is a useful quantity of a search strategy if one
is not interested in how long the search might take. However it does not
provide any information about the efficiency of a search process in terms
of average times for target location. Such information is crucial to
assess real world search scenarios \cite{special} occurring over a wide range of
spatio-temporal scales, from the search of transcription factor proteins for a
specific place on
a DNA chain \cite{bvh,mirny2007,max2013,ottoPRL,PhysRep2012,kolomeisky}
 over food search by animals \cite{Intermittent,viswanathan,mendez2014}
up to rescue operations \cite{rescue} or algorithms for finding the minima
in a complex search space \cite{pavlyukevich}. Any good measure of search
 efficiency must take into account the specific nature of a respective
search process, which a searcher may seek to optimise. 

Brownian motion was considered to be the default for a successful random
search strategy in most cases until the 1980ies. In 1986 Shlesinger and
 Klafter challenged this dogma suggesting that L\'evy flights (LFs) represent
a better strategy if a searcher looks for sparsely distributed targets
\cite{shlesinger}. Due to the divergence of the mean squared average of the
jump length $\ell$ the trajectory of LFs has a different fractal dimension
than the trajectory of Brownian motion \cite{hughes,review}.  This allows
one to avoid oversampling, i.e., revisiting the same point several times,
which is typical for recurrent Brownian motion in one and two dimensions
\cite{viswanathan,viswanathan2002}. In the 1990's L\'evy motion was put
forward as an optimal foraging strategy for animals searching for sparse
food \cite{viswanathan,viswanathan1996,viswanathan1999} and the approach
was recently extended to patchy environments \cite{viswanathan2015}.
This so-called LF hypothesis triggered a vivid debate
 \cite{edwards2007,Benh99,PMC13} in the field of movement ecology
\cite{viswanathan,mendez2014,nathan2008}. It was argued that L\'evy-like
motion has been observed for many animals such as albatrosses \cite{sims2012},
marine predators \cite{sims2008,humphries2010}, terrestrial animals like goats
and deer \cite{goat,deer} and even microzooplankton \cite{dinoflagellate}.
Heavy-tailed distributions were also reported to characterise human movement
patterns \cite{brockmann1,brockmann2}.

\begin{figure}\center 
\includegraphics[height=4.5cm]{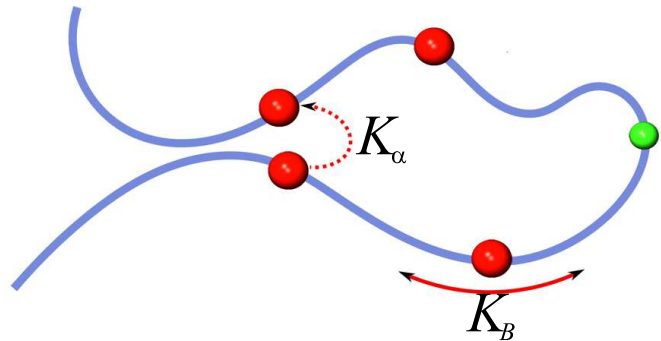}
\caption{Target search of a protein, or enzyme, along a fast-folding DNA
chain allowing a dimensional reduction modelled by intermittent motion:
$K_{\alpha}$ and $K_{B}$ denote the switching rates for performing LFs
representing intersegmental transfers through the bulk due to
unbinding and rebinding (arrows), respectively Brownian motion mimicking
1D Brownian sliding along the polymer chain until the moving particle (red)
finds the target (green) \cite{prl2005}.}
\label{sketch2}
\end{figure}

However, for certain animals the visual perception of their environment
becomes more limited when moving with higher velocity. Observations show
that, to remedy this, in these cases the search process alternates between
a slow recognition mode during which a target can be found, and fast
relocation events where the searcher is insensitive to any target search
\cite{Intermittent,AmSci1990,AmerZool2001}, see Fig. \ref{sketch1}. This
poses the need for theoretical modelling to combine saltatory, jump-like, with
cruise motion yielding {\em intermittent strategies}, which feature combinations
of at least two different types of motion e.g., Brownian and ballistic
motion, or Brownian motion and LFs \cite{Intermittent,gleb1,gleb2}. A
related type of intermittent dynamics is that of composite Brownian motion
\cite{Benh99,PMC13}, which was proposed to model the search of a forager or
particle in patchy environments. Here inter and intra patch movements are
defined by a combination of Brownian modes with different mean step lengths,
however, the searcher can detect targets in both modes. Recently it was
argued that this type of dynamics was observed in the movements of mussels
\cite{mashanova1,mashanova2}.  Composite Brownian motion can be generalised
to an adaptive L\'evy walk, where the Brownian inter-patch movement is
replaced by L\'evy motion \cite{ReynoldsPhysA2009}. Intermittent dynamics
consisting of Brownian and L\'{e}vy motion has indeed been observed for a
number of biological organisms, such as microzooplankton depending on the
density of the prey \cite{dinoflagellate}, coastal jellyfish \cite{HBDF12},
mussels moving in dense environments \cite{dJBKW13} and a variety of marine
predators hunting in different environments \cite{Sims10,SHBB12}. Further
 generalisations of such models use switching rates \cite{prl2005},
or sample the switching times from one mode to another from different
distributions \cite{LomholtPNAS2008}.  The latter type of modelling was
 motivated by studying the target search of proteins on fast-folding
polymer chains, see Fig. \ref{sketch2}. Formally, these models form special
cases of distributed order fractional diffusion equations \cite{CSK11}
or diffusion equations whose Laplacian is augmented with a space fractional
term \cite{prl2005}. They have also been derived as long-time approximations
for correlated L\'evy walks \cite{TKKFG16}. An optimal strategy---in the
sense of maximising a chosen efficiency---thus depends at least on the
type of motion, the switching distributions and the dimension of the search
space \cite{Intermittent,LomholtPNAS2008}. On a molecular scale the search
patterns of regulatory proteins for their target binding site on finite DNA
chains is an LF with a cutoff for the jump length distribution, however,
the advantage of the combined search modes through the bulk and along the
DNA significantly improve the search rate \cite{PNAS2009,PNAS2008,MaxSciRep}. 

In previous works the search reliability and efficiency for LFs
and Brownian motion were studied separately and compared with
each other \cite{PNAS14,LevyLong}. Motivated by the examples of
intermittent motion referred to above, we here
examine a combined process consisting of both Brownian and L\'evy
components. Our paper is structured as follows: after defining the
quantities of interest (the search efficiency and reliability) in section 2,
section 3 recalls the results for pure Brownian and LF search. Sections 4 \&
5, respectively, then present our results for the search reliability and
search efficiency. In section 6 we provide a discussion of our results.
Details for the analytical calculations are provided in the Appendix.

\section{Quantities of interest}

We characterise a search strategy by two different quantities. The first
one is the search reliability, which is the cumulative probability $P$ of
ever reaching the target.  It can be expressed through the Laplace image
of  first arrival time density$\wp_{\mathrm{fa}}(t)$ as \cite{PNAS14}
\begin{equation}
P=\lim_{s\rightarrow0}\wp_{\mathrm{fa}}(s),
\label{Pdef}
\end{equation}
where the Laplace transform of a function $f(t)$ is defined via
$f(s)=\int_0^\infty f(t)e^{-st}dt$. The search reliability depends on the
type of random walk as well as geometrical details (dimension, distance from
the starting position to the target etc.).
Thus $P=1-\mathscr{S}$, where $\mathscr{S}$ is the survival
probability \cite{redner,OshaninMetzlerRedner}. The latter quantity can
be tackled by solving a (fractional) Fokker-Planck equation with a sink
term \cite{PNAS14,LevyLong,JPA2003}. For search in one dimension by LFs
without a bias the search reliability is unity if $\alpha>1$
and zero otherwise \cite{JPA2003}, which is consistent with previous
results \cite{spitzer}. For search in the presence of a bias the search
reliability can vary between zero and unity \cite{PNAS14,LevyLong}, which
is true even for Brownian motion \cite{redner}, where for the case of the
bias pushing a searcher from the target the search reliability  is described
by an exponential (Boltzmann) factor \cite{redner}. A search reliability of
unity does not necessarily imply recurrence of the motion.  For instance,
LFs with $\alpha=1$ in one dimensional and Brownian motion in two dimensions
are recurrent but the search reliability is 0.

The second quantity of interest is the search efficiency. Most of the
theoretical studies consider a probabilistic searcher with a limited radius of
perception. Motivated by \cite{james2010}, in this case two basic definitions
of the search efficiency are considered to be either
\begin{equation}
\mathrm{Efficiency}_1=\frac{\mathrm{visited\,number\,of\,targets}}
{\mathrm{number\,of\,steps}},
\label{eff1}
\end{equation}
or
\begin{equation}
\mathrm{Efficiency}_2=\frac{\mathrm{visited\,number\,of\,targets}}{\mathrm{distance\,travelled}}.
\label{eff2}
\end{equation}
The first definition applies especially to {\em saltatory search}, where
a searcher moves in a jump-like fashion and is able to detect the target
only around the landing point after a jump. The second formula is adapted
to {\em cruise motion}, where the searcher keeps exploring the search space
continuously during the whole search process. An example for the former
scenario  is given by a regulatory protein that moves in three dimensional
space and occasionally binds to the DNA of a biological cell until it finds
its binding \cite{mirny2007,max2013,ottoPRL,PhysRep2012,kolomeisky}. The
latter scenario would correspond to an eagle or vulture whose excellent
eyesight permits them to scan their environment for food during their entire
flight.  For LFs Eq. (\ref{eff1}) presents a natural choice while
Eq. (\ref{eff2}) is better suited for processes like Brownian motion and
L\'evy walks \cite{ZaburdaevReview}.

In this paper we focus on the limit of a sparse target density, which is
approximated by the situation when only one target can be found. For a
single target and saltatory motion we argued that the efficiency should be
defined from Eq. (\ref{eff1}) with a proper averaging \cite{LevyLong}. In
our continuous time model the number of steps from (\ref{eff1}) is naturally
substituted by the time of the process. Since we have one target, the number
of targets found on average can be less than one. Obviously, a time averaging
is needed, and, hence, we choose
\begin{equation}
\mathcal{E}=\left<\frac{1}{t}\right>=\int^\infty_0 \wp_\mathrm{fa}(s)ds.
\label{Edef}
\end{equation}
Below we use the search reliability \& efficiency in the sense of
Eqs.~(\ref{Pdef}) \& (\ref{Edef}) to characterise search strategy of combined
L\'evy-Brownian motion.

\section{First arrival density from a fractional Fokker-Planck equation}

The search properties of a process  combining LFs and Brownian motion
can be effectively calculated from a space-fractional Fokker-Planck diffusion
equation similar to the one considered in \cite{prl2005,JPA2003} for
the non-normalised probability density function (PDF) $f(x,t)$,
\begin{equation}
\frac{\partial f(x,t)}{\partial t}=K_{\alpha}\frac{\partial^{\alpha}f(x,t)}{
\partial \left\vert x \right\vert^{\alpha}}+K_{B}\frac{\partial^{2}f(x,t)}{
\partial x^{2}}-\wp_{\mathrm{fa}}(t)\delta(x),
\label{SinkPolymerFFPE}
\end{equation}
where without losing generality the target is located at $x=0$. We assume that
at $t=0$ the searcher is placed at $x=x_0$, i.e., $f(x,0)=\delta(x-x_0)$. The
consequence of the $\delta$-sink at $x=0$ is the condition $f(0,t)=0$
\cite{prl2005,JPA2003}. The fractional derivative $\partial^\alpha/\partial
x^\alpha$ can be introduced in terms of its Fourier transform,
\begin{equation}
\int^{\infty}_{-\infty}e^{ikx}\left[\frac{\partial^\alpha}{\partial x^\alpha}
f(x,t)\right]dx=-|k|^\alpha f(k,t).
\end{equation} 
We should note here that in Ref. \cite{prl2005} a similar but more specific
equation was used for the description of the problem of protein diffusion on a
polymer chain. In comparison with our Eq. (\ref{SinkPolymerFFPE}) it contained
 additionally two terms, which described the contributions from an adsorption
and desorption of particles modelling the exchange of particles with the
ambient bulk solvent, and was solved for different initial conditions. For
this specific problem the optimal search minimised the mean first arrival
time, which is always finite for that case. In our case this quantity can
become infinitely large, hence the analysis in Ref. \cite{prl2005} is not
applicable to the physical situation considered here.

Integration over the position coordinate of Eq.~(\ref{SinkPolymerFFPE}) yields
\begin{equation}
\wp_{\mathrm{fa}}(t)=-\frac{d}{dt}\int_{-\infty}^{\infty} f(x,t)dx.
\end{equation}
Hence $\wp_{\mathrm{fa}}(t)$ is the negative time derivative of the
survival probability, i.e. $\wp_{\mathrm{fa}}(t)$ is indeed the probability
of first arrival: as soon as a walker gets to the sink it is absorbed.\footnote{We
remind the reader that for Brownian motion first arrival and first passage
lead to identical results, whereas both definitions are conceptually different
for LFs \cite{JPA2003,tal}.}

Analogously to pure search by LFs \cite{JPA2003} it is easy to find
a solution $f(k,s)$ of Eq. (\ref{SinkPolymerFFPE}) in Fourier-Laplace
space
\begin{equation}
f(k,s)=\frac{e^{ikx_0}-\wp_{\mathrm{fa}}(s,x_0)}{s+K_\alpha\left\vert k \right\vert^{\alpha}+K_Bk^2},
\label{fks}
\end{equation}
see also \cite{PNAS14}. Integration of Eq. (\ref{fks}) over $k$ yields: 
\begin{eqnarray}
\int_{-\infty}^{\infty} f(k,s)dk=f(x=0,s)=0=W(-x_0,s)-W(0,s)\wp_{\mathrm{fa}}
(s,x_0),
\end{eqnarray}
where $W(x,s)$ is a solution of Eq. (\ref{SinkPolymerFFPE}) without the sink term.
Hence the probability of first arrival becomes
\begin{equation}
\wp_{\mathrm{fa}}(s)=\frac{\displaystyle\int_{-\infty}^{\infty}dk\frac{e^{ikx_0}}{
s+K_\alpha\left\vert k \right\vert^{\alpha}+K_Bk^2}}{\displaystyle\int_{-\infty}^{
\infty}dk\frac{1}{s+K_\alpha\left\vert k \right\vert^{\alpha}+K_Bk^2}}.
\label{pfa}
\end{equation}
Equation~(\ref{pfa}) can be expressed as a function of dimensionless
variables as:
\begin{eqnarray}
\wp_{\mathrm{fa}}(s)=\frac{\displaystyle\int_0^{\infty}\frac{\cos k}{st_B+p
k^{\alpha}+k^2}dk}{\displaystyle\int_0^{\infty}\frac{1}{st_B+p k^{\alpha}+k^2}dk},
\label{pfp}
\end{eqnarray}
where $t_B=x_0^2/K_B$ is the time scale set by Brownian motion over the length
$x_0$ and
\begin{equation}
p=x_0^{2-\alpha}K_\alpha/K_B.
\label{ppar}
\end{equation}
According to Eqs.~(\ref{Pdef}) and (\ref{Edef}), in order to get the
reliability and efficiency of the combined search, one has to compute,
respectively, the limit $s\rightarrow0$ of Eq.~(\ref{pfp}) and the integral
of Eq.~(\ref{pfp}) over $s$ from zero to infinity.  Before going into this
analysis and its consequences, we recall the results for search by Brownian
motion and LFs strategies separately.

\subsection{Brownian search}

If the search process proceeds only with Brownian moves, i.e., $K_\alpha=0$,
the first arrival density can be computed analytically and in Laplace space
reads
\begin{equation}
\wp_{\mathrm{fa}}(s)=\exp\left(-x_0\sqrt{\frac{s}{K_{B}}}\right)
\label{pfabrownbias}
\end{equation}  
or, back-transformed to time, 
\begin{equation}
\wp_{\mathrm{fa}}(t)=\frac{x_0}{\sqrt{4\pi K_{B}t^3}}\exp\left(-\frac{x_0^2}{
4K_{B}t}\right).
\label{alpha2}
\end{equation}
This is well known L\'evy-Smirnov density \cite{OshaninMetzlerRedner}. Obviously
in this case the search reliability (\ref{Pdef}) is $P=1$, and the efficiency
$\mathcal{E}_B=\frac{2K_B}{x_0^2}$.

\subsection{First arrival for pure L\'evy search}

The expression for $\wp_{\mathrm{fa}}(s)$ for pure L\'evy search can be
computed in terms of Fox $H$-functions \cite{LevyLong}. In the limit of
small $s$ corresponding to the long time limit, $\wp_{\mathrm{fa}}(s)$ can
be computed in terms of elementary functions (see \ref{levypurelongt}). For
$\alpha\le1$ the search reliability is $P=0$, for $\alpha>1$ the reliability is
$P=1$. By integration of the corresponding $H$-function expression in Laplace
space one gets the simple equation for the search efficiency \cite{LevyLong}
(see also a derivation without use of $H$-functions in \ref{PureLevyEff}),
\begin{eqnarray}
\mathcal{E}_\alpha=\left\langle\frac{1}{t}\right\rangle=\frac{\alpha K_\alpha}{
x_0^\alpha}\cos\left(\pi\left(1-\frac{\alpha}{2}\right)\right)\Gamma(\alpha),
\qquad1<\alpha<2.
\label{EffLevyFlat}
\end{eqnarray}

\section{Search reliability for combined L\'evy-Brownian search}

In order to compute the search reliability one has to take the limit $s\rightarrow0$ of expression (\ref{pfp}) as pointed out in Eq. (\ref{Pdef}). If $\alpha\geq1$, both integrals in (\ref{pfp}) diverge at $s=0$. The divergence occurs at $k\rightarrow0$, and only the values of the integral very close to $k=0$ (where $\cos k\approx 1$) make a contribution to the integral. Hence $P=1$, which is understandable intuitively, because if one combines two processes with search reliability $P=1$, then the combined process should also have
this property. Interestingly the search reliability for a combination
of LFs with $\alpha=1$ and Brownian motion also has $P=1$,
while pure LF search with $\alpha=1$ is absolutely
unreliable ($P=0$) \cite{PNAS14}.

The case $\alpha<1$ is less trivial. Even if $s=0$, both integrals in
Eq.~(\ref{pfp}) are convergent and the search reliability can be
computed in terms of $H$-functions (\ref{Hsolution}), resulting in
\begin{eqnarray}
P=\frac{\sin\left(\frac{\pi}{2-\alpha}\right)}{2\sqrt\pi}H^{12}_{31}\left[\frac{
2}{p^{1/(2-\alpha)}}\begin{array}{|lll}\left(1,\frac{1}{2}\right)\left(\frac{1-
\alpha}{2-\alpha},\frac{1}{2-\alpha}\right)\left(\frac{1}{2},\frac{1}{2}\right)\\
\left(\frac{1-\alpha}{2-\alpha},\frac{1}{2-\alpha}\right) \end{array}\right],
\label{Palpha}
\end{eqnarray}
where the parameter $p$ is defined in Eq.~(\ref{ppar}). Result (\ref{Palpha})
is naturally independent of the time $t_B$, because rescaling of the time
should not change the search reliability.  In the case $\alpha=0$  the
corresponding fractional derivative in Eq. (\ref{SinkPolymerFFPE}) is of
zeroth order, which corresponds to a Fokker-Planck equation with decay term
$-f(x,t)$. The result (\ref{Palpha}) simplifies to $P(\alpha=0)=\exp(-\sqrt
p)=\exp(-\sqrt{x_0^{2-\alpha}K_\alpha/K_B})$ in complete agreement with the
solution of diffusion equation with decay term.

The standard expansion for $H$-functions \cite{prudnikov} can be used to
find the leading behaviour of Eq.~(\ref{Palpha}) in the limit of small $p$,
\begin{equation}
P(p\ll1)\approx1-C_1(\alpha)p^{\frac{1}{2-\alpha}}+\frac{1}{2}p^{\frac{2}{2-
\alpha}}-C_2(\alpha)p^{\frac{3-\alpha}{2-\alpha}},
\label{Pexp}
\end{equation}
where the coefficients are defined as\footnote{In the limit of $\alpha\to1$ the
coefficient $C_1(\alpha)$ vanishes. However $P$ does not exceed the value unity,
as in this limit we should include the third term in the expansion, as it has the
same power. Then $C_3$ exactly cancels with $C_2$ and no contradiction appears.}
\begin{eqnarray}
\nonumber
&C_1(\alpha)=\frac{2-\alpha}{2}\sin\left(\frac{\pi}{2-\alpha}\right),\\
&C_2(\alpha )=\frac{(2-\alpha)\sin\left(\frac{\pi}{2-\alpha}\right)}{2^{4-
\alpha}\sqrt\pi}\frac{\Gamma\left(-\frac{3}{2}+\frac{\alpha}{2}\right)}{
\Gamma\left(2-\frac{\alpha}{2}\right)}.
\end{eqnarray}

\begin{figure}\center
\includegraphics[width=12cm]{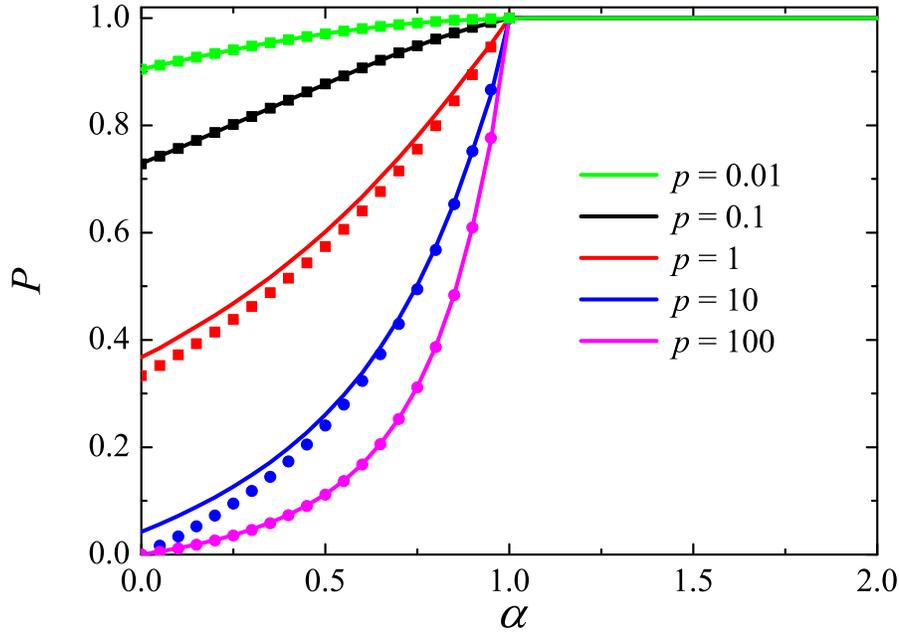}
\caption{Search reliability P as function of the L\'evy stable exponent
$\alpha$ for various values of $p$, Eq.~(\ref{ppar}). The continuous curves
are obtained by numerical computation of Eq. (\ref{pfp}). The squares for the
cases $p=0.01$ (green), $p=0.1$ (black) and $p=1$ (red) are obtained from the
approximation (\ref{Pexp}) for small $p$. The circles of corresponding colour
are plotted from the approximate expression (\ref{Ppgg1}) for large $p$.}
\label{SearchRelAlpha}
\end{figure}

From Eq. (\ref{pfp}) one can also find an expansion for $P$ in the limit of
large $p$, which reads (for the derivation see \ref{pbig}),
\begin{equation}
P(p\gg1)\approx\frac{1}{\pi}\Gamma(1-\alpha)\sin\left(\frac{\pi\alpha}{2}\right)
(2-\alpha)\sin\left(\frac{\pi}{2-\alpha}\right) p^{(\alpha-1)/(2-\alpha)}.
\label{Ppgg1} 
\end{equation}
Note that the limiting value of the latter expression for $\alpha\rightarrow1-$
is 1, i.e., the divergence of $\Gamma(1-\alpha)$ is exactly compensated by the
convergence of $\sin(\pi/(2-\alpha))$ to zero.

In Fig. 3 the search reliability $P$ is plotted as a function of the stable
index $\alpha$.  As discussed above, for $\alpha\ge1$ the value of $P$ is
unity, i.e. the combined L\'evy-Brownian search is absolutely reliable.
However, when $\alpha<1$ the hitting probability is less than unity and
 decreases with $\alpha$ until it reaches the values for the diffusion
equation with decay term. The curves were obtained from the numerical
computation of the integral ratio in Eq. (\ref{pfp}). The expansion
of Eq. (\ref{Palpha}) in the limit of small $p$, i.e. Eq. (\ref{Pexp})
gives a very good approximation for $p\lesssim1$ (green, red and black
squares). For $p\gtrsim10$ the expansion Eq. (\ref{Ppgg1}) for the limit
$p\gg1$ works quite well even for small $\alpha$. For $\alpha$ close to
unity, it approximates the numerical solution nicely even for $p\gtrsim1$.

\begin{figure}\center
\includegraphics[width=12cm]{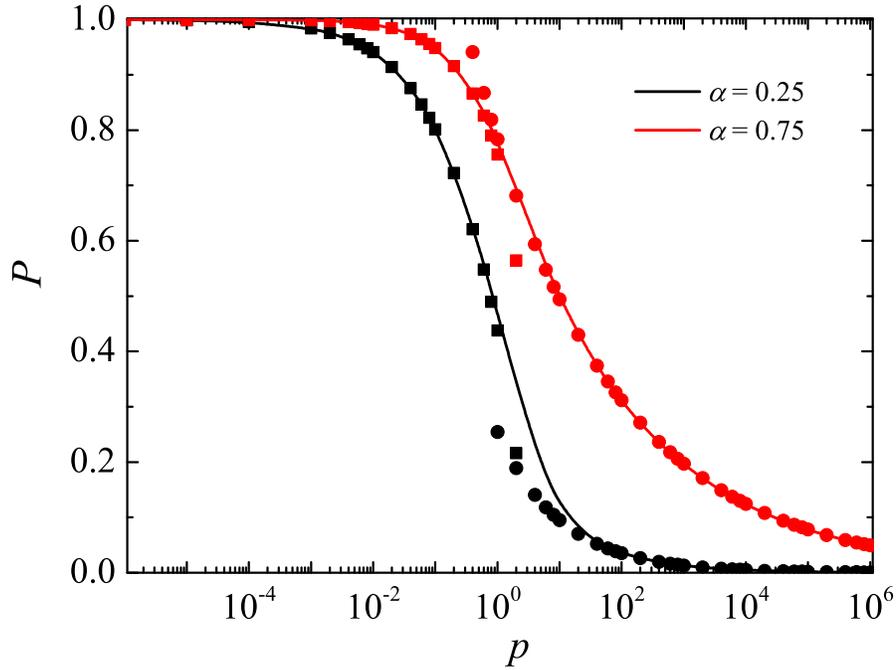}
\caption{Search reliability as a function of the parameter $p=x_0^{2-\alpha}
K_\alpha/K_B$ for various stable indices $\alpha$. Continuous lines were
obtained from numerical solution of Eq. (\ref{pfp}). The squares show the
analytical approximation (\ref{Pexp}) for small $p$. The circles are
plotted from the approximate expression (\ref{Ppgg1}) for large $p$.}
\label{SearchRelx0}
\end{figure}

Obviously $P$ should depend on the distance between the starting position and the target through the parameter $p$ defined in Eq.~(\ref{ppar}). If in the case of Figs. \ref{SearchRelAlpha} and \ref{SearchRelx0} one fixes the diffusion coefficients, for instance, $K_\alpha=1 \frac{cm^\alpha}{sec}$ and $K_B=1\frac{cm^2}{sec}$, then the dependence on $p$ is essentially a dependence on the initial distance between the searcher and the target to the power $2-\alpha$. In Fig. \ref{SearchRelAlpha} this dependence is displayed in the order of the curves: the higher the value of $p$ the lower is the curve in the plot. In Fig. \ref{SearchRelx0}
this dependence is shown explicitly. The larger the initial separation
$x_0$ between the searcher and the target the smaller becomes the value of
$P$. The decrease in $\alpha$, which corresponds to the higher fraction
of long jumps, leads to a drop in the search reliability.  Circles show
the large $p$ approximation (\ref{Ppgg1}), while squares correspond to the
small $p$ formula (\ref{Pexp}). One can see that for $\alpha=0.75$ (red
symbols/curve) these two limiting expressions describe the whole curve quite
well. Even for the quite low value $\alpha=0.25$ (black symbols/curve) the
quality of the correspondence of the asymptotic formulas is still very good.

\section{Search efficiency for combined L\'evy-Brownian search}

\begin{figure}\center
\includegraphics[width=16cm]{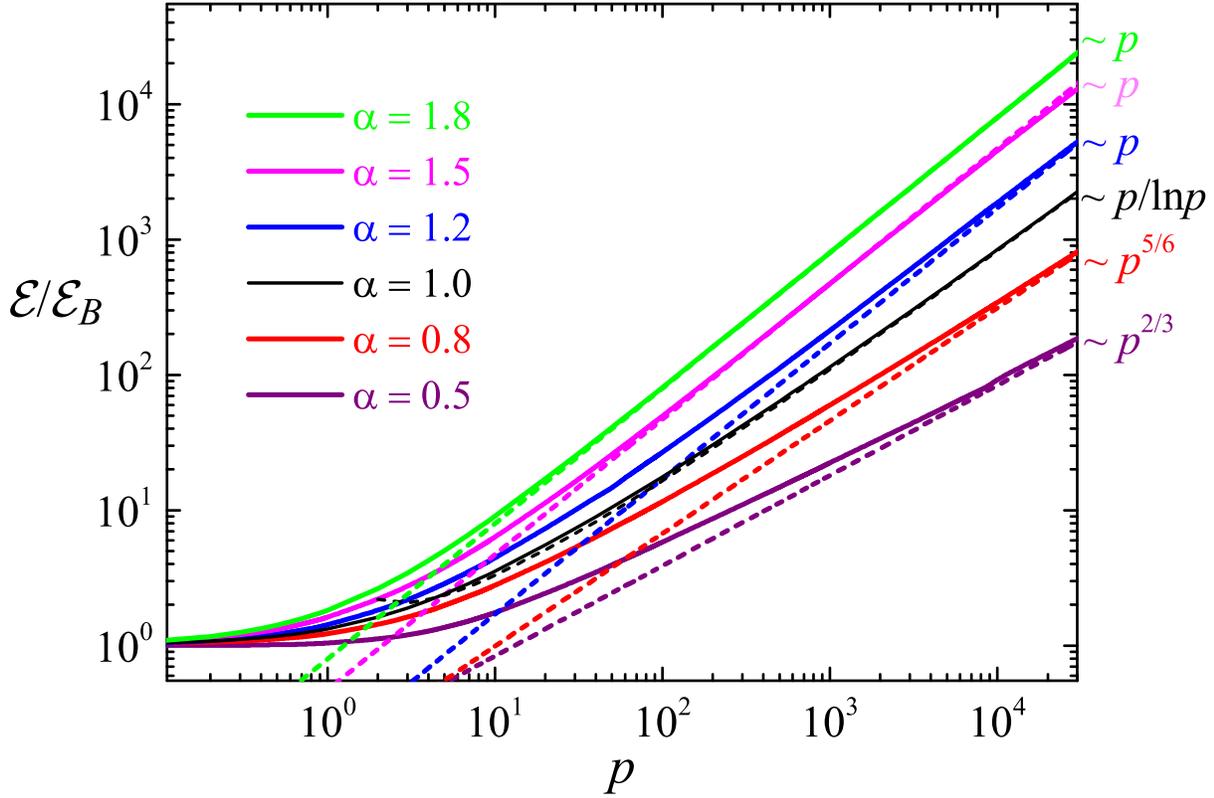}
\caption{Search efficiency as function of the parameter $p=x_0^{2-\alpha}
K_\alpha/K_B$. Continuous curves correspond to numerical results for the
efficiency of the Brown-LF strategy for various values of the stable
index $\alpha$. Dashed lines are asymptotics for the limit $p\rightarrow
\infty$, which are derived in \ref{asymptotic}.}
\label{EffP}
\end{figure}

One can rewrite the definition of the search efficiency (4) as follows,
\begin{eqnarray}
\mathcal{E}\left(p\right)=\frac{\mathcal{E}_B}{2}\int_0^\infty\wp_{\mathrm{fa}}
(p,s)d(st_{B}),
\label{effdefnorm}
\end{eqnarray}
where $\mathcal{E}_B$ is the Brownian efficiency $\mathcal{E}_B=2K_B/x_0^2$
of the search process with $p=0$ and $\wp_{\mathrm{fa}}(p,s)$
is determined by Eq. (\ref{pfp}). If $p=0$, then $\int_0^\infty
\wp_{\mathrm{fa}}(p,s)d(st_{B})=2$, i.e. $\mathcal{E}(p=0)=\mathcal{E_B}$, as
it should be. The value of the integral in Eq. (\ref{effdefnorm}) does not
depend on the time $t_B$ but only on the value of the parameter $p$. Hence
in Fig.~\ref{EffP} we plot the ratio of the efficiency of combined Brown-LF
search normalised by the efficiency $\mathcal{E}_B$ of pure Brownian search as
a function of the parameter $p$. The continuous curves represent numerical
results, and the dashed lines are asymptotes in the limit of large $p$
(for the derivation see \ref{asymptotic}). First, we see that in all cases
the efficiency increases monotonically with $p$. If $p$ is constant then
the decrease of $\alpha$ leads to a monotonic decrease of the efficiency.
Secondly, there is a qualitative difference for the Brownian-L\'evy search
with $\alpha>1$ and $\alpha<1$. In the former case (for which, as we know,
LFs without Brownian motion find the target anyway) the efficiency in the
limit of large $p$ is proportional to $p$. A careful calculation shows
that in this case $\mathcal{E}=\mathcal{E_\alpha}$ (\ref{asymptotic}),
i.e., the search efficiency is determined only by LFs. This is reasonable,
because if both processes lead to the location of the target, only the one
with a very large noise strength should matter. In the latter case for
which pure LFs would not succeed ($\mathcal{E}_\alpha=0$) the asymptotic
dependence is of power law form $p^{1/(2-\alpha)}$, i.e. even the smallest
fraction of Brownian motion makes the location of the target possible. This
 observation is consistent with the fact that the search reliability does
not fall off to zero at $\alpha\le1$ (cf. Fig. \ref{SearchRelAlpha}) The
limiting case $\alpha=1$ shows a logarithmic correction. An interesting
point here is that for $\alpha=1$ pure LF search is absolutely unreliable,
$P=0$, but the smallest contribution of Brownian motion makes the search
absolutely reliable, $P=1$.

Although Fig.~\ref{EffP} shows the complete dependence of the efficiency
on all original parameters $K_\alpha$, $K_B$, $x_0$, and $\alpha$, which
are combined into a single parameter $p$, the plot should be interpreted
carefully. At first glance it seems that an increase of $\alpha$ leads to
an increase of the search efficiency if $p=const$, i.e., Brownian motion
will be the most efficient search strategy. However, any change of $\alpha$
implicitly affects the parameter $p$, compare Eq.~(\ref{ppar}). Even if
one assumes that $K_B$ is constant the fractional diffusion coefficient
$K_\alpha$ will change its dimension with change of $\alpha$ and in order to
keep a fixed value of $p$ one needs to change the distance from the target.

\begin{figure}\center
\includegraphics[width=11cm]{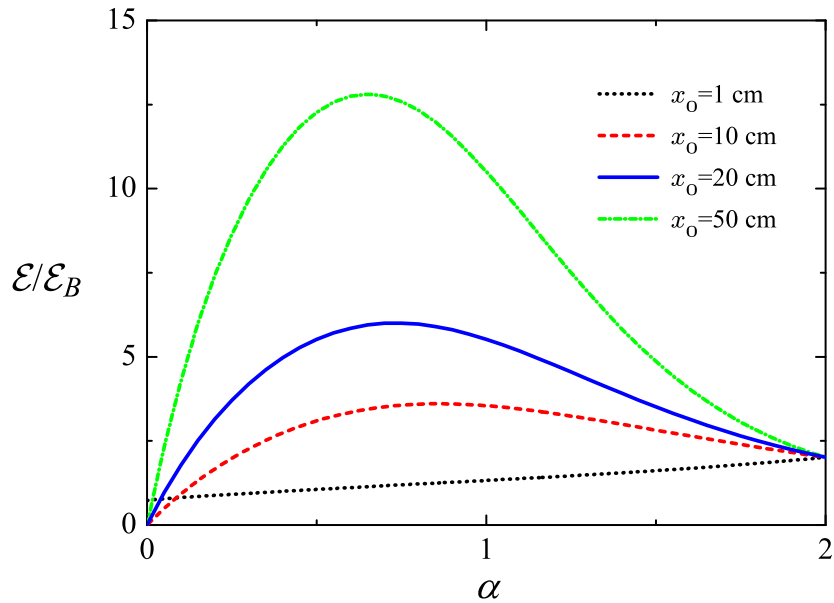}
\caption{Search efficiency as function of the stable index $\alpha$ for
fixed $x_0$, $K_\alpha=1\, \mathrm{cm^\alpha/sec},K_2=1\, \mathrm{cm^2/sec}$.}
\label{Effx0fixed}
\end{figure}

In order to fix the starting position and compare the strategies
in this practically important case we plot the search efficiency
as a function of $\alpha$ for fixed $x_0=1$, 10, 20, 50 cm and
$K_\alpha=1 \mathrm{cm^\alpha/sec}$, $K_2=1\mathrm{cm^2/sec}$ in
Fig.~\ref{Effx0fixed}. The curve for $x_0=1$ cm monotonically increases
with $\alpha$, i.e., the Brownian strategy is optimal for finding a nearby
target. However, for larger values of $x_0$ a maximum appears, which shows
that the combination of Brownian motion with LFs may perform better for
larger initial separation from the target similar to the case of pure LF
search as discussed in Ref.~\cite{PNAS14}.

\section{Discussion and conclusions}

To summarise, we found that for complex motion which combines LFs for
$\alpha<1$ with Brownian motion, the search reliability can have intermediate
values between zero and unity even if no bias is present. If the L\'evy stable
exponent $\alpha$ is larger or equal than unity, then $P=1$. For a process
which combines two L\'evy motions the qualitative behaviour is the same if one
of the exponents is larger than unity. If both of them are smaller than unity,
then the process is transient, $P=0$, and a point-like searcher is unable
to find a point-like target. It is interesting to compare our findings with
those for the pure L\'evy search for a delocalised target---with power-law
absorption $a(x)=1/(\vert x\vert^{\beta}+1)$, i.e. the target is discovered
with a power-law decaying probability---in Ref. \cite{elena2015}. In that case,
if the stable index $\alpha<1,$ the search still can be absolutely reliable
($P=1$), if the scaling exponent characterising an absorption probability
$\beta\le\alpha$, i.e., the target localisation or absorption probability
is delocalised stronger than the LF process. 

The search efficiency has a universal behaviour as function of the 
dimensionless parameter $p$, which describes the ratio of the noise
intensities of the different modes. If the characteristic exponent of the
LFs is larger than unity---which includes also Brownian motion---then for
large $p$ the efficiency is linear in $p$ reflecting the fact that only
one of the modes defines the properties of the trajectories. Once $\alpha$
becomes unity a logarithmic correction factor appears and the efficiency
grows sublinear with $p$. If the L\'evy exponent $\alpha$ is less than
unity, the efficiency grows with $p$ as a sublinear power law. The latter
case shows that even for high intensity of LF search with $\alpha<1$
the local search provided by Brownian motion matters and quantitatively
affects the search efficiency.

The optimisation of a combined search strategy of LFs with stable index
$\alpha$ and Brownian motion shows that for targets in the close vicinity of
the starting point one should use more local search strategies, i.e., $\alpha$
should be close to 2, while for distant targets a larger fraction of long
jumps increases the search efficiency. This result is consistent with the
one-mode case in which LFs are confronted with Brownian motion \cite{PNAS14}.

It would be interesting to apply our theory to better understand the
combined dynamics of biological organisms such as microzooplankton
\cite{dinoflagellate}, coastal jellyfish \cite{HBDF12}, moving mussels
\cite{dJBKW13} and marine predators hunting in different environments
\cite{SHBB12,Sims10}.

\appendix

\section{Long time limit of the first arrival density and the search reliability
for pure LF search}
\label{levypurelongt}

The probability density of the first arrival for pure LF search is
\begin{equation}
\wp_{\mathrm{fa}}(s)=\frac{\displaystyle\int_0^{\infty}\frac{\cos kx_0}{s+
K_\alpha k^{\alpha}}dk}{\displaystyle\int_0^{\infty}\frac{1}{s+K_\alpha
k^{\alpha}}dk}.
\label{pfapurelevy}
\end{equation}
In the case $\alpha<1$ for $s=0$ the numerator is finite while the denominator
diverges for any $s$, i.e., the search reliability becomes $P=0$. For $\alpha
= 1$ the numerator is finite for $s\neq0$, whereas the denominator still
diverges for all $s$, thus we have $P=0$ again. In the case $\alpha>1$
both the numerator and denominator converge at finite $s$. Thus, the search
reliability is non-zero. Let us consider the case of small $s$ corresponding
to long times. The expression (\ref{pfapurelevy}) can be transformed as
\begin{eqnarray}
\nonumber
\wp_{\mathrm{fa}}(s)&=&\frac{\displaystyle\int_0^{\infty}\frac{1}{s+K_\alpha
k^{\alpha}}dk-\int_0^{\infty}\frac{1-\cos (kx_0)}{s+K_\alpha
k^{\alpha}}dk}{\displaystyle\int_0^{\infty}\frac{1}{s+K_\alpha
k^{\alpha}}dk}\\
&=&1-\frac{\displaystyle\int_0^{\infty}\frac{1-\cos(kx_0)}{s+K_\alpha
k^{\alpha}}dk}{\displaystyle\int_0^{\infty}\frac{1}{s+K_\alpha k^{\alpha}}dk}.
\end{eqnarray}
The integral in the denominator can be computed easily (Ref. \cite{prudnikov},
(2.2.3.5)),
\begin{equation}
\int_0^{\infty}\frac{1}{s+K_\alpha k^{\alpha}}dk=\frac{1}{K_\alpha^{1/\alpha}}
\int_0^{\infty}\frac{1}{s+y^{\alpha}}dy=\frac{\pi}{\alpha\sin\left(\pi/\alpha
\right)}\frac{s^{1/\alpha-1}}{K_\alpha^{1/\alpha}}.
\end{equation}
Since the integral in the numerator converges at $s\rightarrow0$, we can simply
put $s=0$ while looking at small $s$ (long time) behaviour. Thus we have
\begin{eqnarray}
\nonumber
\int_0^{\infty}\frac{1-\cos(kx_{0})}{s+K_\alpha k^{\alpha}}dk&=&\frac{x_0^{
\alpha-1}}{(\alpha-1)K_\alpha}\int_0^\infty dy\frac{\sin y}{y^{\alpha-1}}\\
&=&\frac{\Gamma(2-\alpha)}{\alpha-1}\sin\left(\frac{\pi\alpha}{2}\right)
\frac{x_0^{\alpha-1}}{K_\alpha}.
\label{PdenomLevy}
\end{eqnarray}
Hence we get:
\begin{equation}
\wp_{\mathrm{fa}}(s)\approx1-\Lambda(\alpha)x_0^{\alpha-1}K_\alpha^{1/\alpha-1}
s^{1-1/\alpha},
\end{equation}
where
\begin{equation}
\Lambda(\alpha)=\frac{\alpha\Gamma(2-\alpha)}{\pi(\alpha-1)}\sin\left(\frac{\pi
\alpha}{2}\right)\sin\left(\frac{\pi}{\alpha}\right).
\end{equation}
We see that in this case $P=\lim_{s\rightarrow0}\wp_{\mathrm{fa}}(s)=1$. In order
to get the long-time limit with a small-$s$ expansion we note that
\begin{equation}
-\wp_{\mathrm{fa}}'(s)\approx\Lambda(\alpha)x_0^{\alpha-1}K_\alpha^{1/\alpha-1}\left(
1-\frac{1}{\alpha}\right)s^{-1/\alpha}.
\end{equation}
The latter expression is the Laplace transform of $t\wp_{\mathrm{fa}}(t)$. Hence,
according to Tauberian theorems
\begin{equation}
t\wp_\mathrm{fa}(t)\approx\Lambda(\alpha)x_0^{\alpha-1}K_\alpha^{1/\alpha-1}\left(1-
\frac{1}{\alpha}\right)\frac{t^{-1+1/\alpha}}{\Gamma(1/\alpha)}.
\end{equation}
Thus,
\begin{equation}
\wp_\mathrm{fa}(t)\approx C(\alpha)x_0^{\alpha-1}K_\alpha^{1/\alpha-1}t^{-2+1/
\alpha},
\end{equation}
where
\begin{equation}
C(\alpha)=\frac{\Gamma(2-\alpha)}{\pi\Gamma\left(1/\alpha\right)}\sin\left(
\frac{\pi\alpha}{2}\right)\sin\left(\frac{\pi}{\alpha}\right).
\end{equation}

\section{Efficiency of pure LF search}
\label{PureLevyEff}

The search efficiency in this case becomes
\begin{equation}
\mathcal{E}=\int_0^\infty ds\wp_{\mathrm{fa}}(s)=K_\alpha\int_0^\infty ds\frac{
\displaystyle\int_0^{\infty}\frac{\cos kx_0}{s+k^{\alpha}}dk}{\displaystyle\int
_0^{\infty}\frac{1}{s+k^{\alpha}}dk}.
\label{Epurelevy}
\end{equation}
The denominator was computed in expression (\ref{PdenomLevy}). We rewrite the
expression for the efficiency using the notation for $\alpha$-stable density
as $l_\alpha(x)$ as well as new variables,
\begin{eqnarray}
\nonumber
\mathcal{E}&=&\frac{K_\alpha\alpha\sin\left(\pi/\alpha\right)}{\pi}\int_0^\infty
ds s^{1-1/\alpha}\int_0^{\infty}\frac{\cos kx_0}{s+k^{\alpha}}dk\\
\nonumber
&=&\frac{K_\alpha\alpha\sin\left(\pi/\alpha\right)}{\pi}\int_0^\infty ds s^{1-1/
\alpha}\int_0^{\infty}\cos (kx_0)dk\int_0^\infty d\tau e^{-(s+k^\alpha)\tau}\\
\nonumber
&=&K_\alpha\alpha\sin\left(\pi/\alpha\right)\int_0^\infty ds s^{1-1/\alpha}\int
_0^\infty d\tau e^{-s\tau}\int_0^{\infty}\frac{\cos (kx_0)}{\pi} e^{-k^{\alpha}
\tau}dk\\
\nonumber
&=&K_\alpha\alpha\sin\left(\pi/\alpha\right)\int_0^\infty d\tau \tau^{-1/\alpha}
l_\alpha\left(\frac{x_0}{\tau^{1/\alpha}}\right)\int_0^\infty ds s^{1-1/\alpha}
e^{-s\tau}\\
\nonumber
&=&K_\alpha\alpha\sin\left(\pi/\alpha\right)\Gamma\left(2-\frac{1}{\alpha}\right)
\int_0^\infty d\tau \tau^{-2}l_\alpha\left(\frac{x_0}{\tau^{1/\alpha}}\right)\\
\nonumber
&=&\left| y=x_0/\tau^{1/\alpha}\right|\\
\nonumber
&=&- K_\alpha\alpha^2\sin\left(\pi/\alpha\right)\Gamma\left(2-\frac{1}{\alpha}
\right)x_0^{-\alpha}\int_0^\infty dy y^{\alpha-1}l_\alpha\left(y\right)\\
&=&\frac{1}{2}K_\alpha\alpha^2\sin\left(-\pi/\alpha\right)\Gamma\left(2-\frac{
1}{\alpha}\right)x_0^{-\alpha}\left\langle \vert y\vert^{\alpha-1} \right\rangle,
\label{B2}
\end{eqnarray}
where $\left\langle \vert y\vert^{\alpha-1} \right\rangle$ is the $(\alpha-1)$-st
moment of a standard $\alpha$-stable distribution, which is given by
\cite{westseshadri82}
\begin{equation}
\langle|y\vert^q|\rangle=\frac{2}{\pi q}\sin\left(\frac{\pi q}{2}\right)\Gamma
\left(1+q\right)\Gamma\left(1-\frac{q}{\alpha}\right),0<q<\alpha.
\end{equation}
After plugging the latter expression in (\ref{B2}) we get the final result
\begin{equation}
\mathcal{E}=\frac{\alpha K_\alpha}{x_0^\alpha}\left\vert\cos\left(\frac{\pi
\alpha}{2}\right)\right\vert\Gamma(\alpha),\qquad 1<\alpha<2.
\label{eff_fin}
\end{equation}

\section{Derivation of the Fox function solution for $\alpha<1$ at $s=0$}
\label{Hsolution}

The search reliability is
\begin{eqnarray}
P=\wp_{\mathrm{fa}}(s=0)=\frac{\displaystyle\int_0^{\infty}\frac{\cos k}{p
k^{\alpha}+k^2}dk}{\displaystyle\int_0^{\infty}\frac{1}{p k^{\alpha}+k^2}dk},
\end{eqnarray}
and we have \cite{prudnikov}
\begin{eqnarray}
\int_0^{\infty}\frac{1}{pk^{\alpha}+k^2}dk=\frac{\pi}{2-\alpha}\frac{p^{-\frac{
1}{2-\alpha}}}{\sin\left(\frac{\pi}{2-\alpha}\right)}.
\label{intergalkalphak2}
\end{eqnarray}
Since
\begin{eqnarray}
\frac{1}{p+k^{2-\alpha}}=\frac{1}{p}\frac{1}{2-\alpha}H^{11}_{11}\left[\frac{k}{
p^{\frac{1}{2-\alpha}}}\left|\begin{array}{l}\left(0,\frac{1}{2-\alpha}\right)\\
\left(0,\frac{1}{2-\alpha}\right)\end{array}\right.\right],
\end{eqnarray}
we can compute the search reliability in terms of a Fox $H$-function
\begin{eqnarray}
\nonumber
\frac{1}{p(2-\alpha)}&\int_0^\infty k^{-\alpha}\cos k H^{11}_{11}\left[\frac{k}
{p^{\frac{1}{2-\alpha}}}\left|\begin{array}{l}\left(0,\frac{1}{2-\alpha}\right)\\
\left(0,\frac{1}{2-\alpha}\right) \end{array}\right.\right]dk\\
&=\frac{\sqrt\pi 2^{-\alpha}}{p(2-\alpha)} H^{12}_{31}\left[\frac{2}{p^\frac{1
}{2-\alpha}}\left|\begin{array}{l}\left(\frac{1+\alpha }{2},\frac{1}{2}\right),
\left(0,\frac{1}{2-\alpha}\right),\left(\frac{\alpha }{2},\frac{1}{2}\right)\\
\left(0,\frac{1}{2-\alpha}\right)\end{array}\right.\right].
\end{eqnarray}
Thus
\begin{eqnarray}
\nonumber
P&=&\frac{2^{-\alpha}\sin\left(\frac{\pi\left(1-\alpha\right)}{2-\alpha}\right)}{
\sqrt\pi p^{\frac{1-\alpha}{2-\alpha}}}H^{12}_{31}\left[\frac{2}{p^\frac{1}{2-
\alpha}}\left|\begin{array}{l}\left(\frac{1+\alpha }{2},\frac{1}{2}\right),\left(
0,\frac{1}{2-\alpha}\right),\left(\frac{\alpha }{2},\frac{1}{2}\right)\\
\left(0,\frac{1}{2-\alpha}\right)\end{array}\right.\right]\\
&=&\frac{\sin\left(\frac{\pi}{2-\alpha}\right)}{2\sqrt\pi}H^{12}_{31}\left[\frac{
2}{p^\frac{1}{2-\alpha}}\left|\begin{array}{l}\left(1,\frac{1}{2}\right),\left(
\frac{1-\alpha}{2-\alpha},\frac{1}{2-\alpha}\right),\left(\frac{1}{2},\frac{1}{2}
\right)\\\left(\frac{1-\alpha}{2-\alpha},\frac{1}{2-\alpha}\right)\end{array}
\right.\right].
\end{eqnarray}
In the important particular case $\alpha=0$,
\begin{eqnarray}
\nonumber
\lim_{\alpha=0}P&=&\frac{1}{2\sqrt\pi}H^{12}_{31}\left[\frac{2}{\sqrt p}\left|
\begin{array}{l}\left(1,\frac{1}{2}\right),\left(\frac{1}{2},\frac{1}{2}\right),
\left(\frac{1}{2},\frac{1}{2}\right)\\
\left(\frac{1}{2},\frac{1}{2}\right)\end{array}\right.\right]\\
\nonumber
&=&\frac{1}{2\sqrt\pi}H^{02}_{20}\left[\frac{2}{\sqrt p}\left|\begin{array}{l}
\left(1,\frac{1}{2}\right),\left(\frac{1}{2},\frac{1}{2}\right)\\
\rule{0.8cm}{0.02cm}\end{array}\right.\right]\\
&=&\exp\left(-\sqrt p\right).
\end{eqnarray}

\section{Search reliability in the $p\gg1$ limit}
\label{pbig}

In order to find the expansion for $P$ in the limit of large values of $p$ we
start from Eq.~(\ref{pfp}), where $s=0$. The upper integral can be expressed as
\begin{equation}
\int_0^{\infty}\frac{\cos kdk}{pk^\alpha+k^2}=\mathrm{Re}\int_0^{\infty}\frac{
e^{ik}dk}{pk^\alpha+k^2}=\mathrm{Re}I,
\end{equation}
where
\begin{equation}
I=\chi^{-1}\int_0^\infty\frac{e^{i\chi\kappa}d\kappa}{\kappa^\alpha+\kappa^2}
=\chi^{-1}\int_0^\infty d\kappa\frac{1}{1+\kappa^{2-\alpha}}\frac{e^{i\chi
\kappa}}{\kappa^\alpha},
\end{equation}
such that
\begin{equation}
\frac{1}{\chi}=\int_0^\infty e^{i\chi\kappa}\kappa^{-\alpha}\phi(\kappa) d\kappa,
\end{equation}
with $\phi(\kappa)=\frac{1}{1+\kappa^{2-\alpha}}$ and
$\chi=p^{\frac{1}{2-\alpha}}$. Since $\chi$ is large, $e^{i\chi k}$ is a
highly oscillating function. In addition, we have an integrated divergence of
the integrand at zero. Due to these two reasons the main contribution to the
integral will be given by the contribution around $\kappa\approx0$. Therefore
(and using the expression 2.3.3.1 from \cite{prudnikov})
\begin{eqnarray}
\nonumber
I&=&\frac{1}{\chi}\int_0^\infty e^{i\chi\kappa}\kappa^{-\alpha}\phi(\kappa)
d\kappa\approx\chi^{-1}\int_0^\infty e^{i\chi\kappa}\kappa^{-\alpha}\phi(0)
d\kappa\\
\nonumber
&=&\chi^{-(2-\alpha)}\int_0^\infty\frac{e^{i\xi}}{\xi^\alpha}d\xi\\
&=&\chi^{-(2-\alpha)}\Gamma(1-\alpha)(-i)^{\alpha-1}=-\frac{\Gamma(1-\alpha)}{p}
e^{-\frac{i\pi}{2}(\alpha-1)}.
\end{eqnarray}
And thus
\begin{equation}
\int_0^{\infty}\frac{\cos kdk}{pk^\alpha+k^2}=\frac{1}{p}\Gamma(1-\alpha)\sin
\left(\frac{\pi\alpha}{2}\right).
\end{equation}
The integral in the denominator of Eq.~(\ref{pfp}) can be computed analytically
for any $p$ and the result is given by Eq.~(\ref{intergalkalphak2}). Hence one
gets Eq.~(\ref{Ppgg1}).

\section{The case $\alpha=1$}
\label{alpha1mu2solution}

For $\alpha=1$,
\begin{equation}
\wp_{\mathrm{fa}}(s)=\frac{\displaystyle\int_{-\infty}^{\infty}dk\frac{e^{ik
x_0}}{s+K_{\alpha}\left|k\right|+K_Bk^2}}{\displaystyle\int_{-\infty}^{\infty}
dk\frac{1}{s+K_{\alpha}\left|k\right|+K_Bk^2}}.
\end{equation}
Obviously the integrals can be simplified due to the symmetry in $k$. Then
\begin{equation}
\wp_{\mathrm{fa}}(s)=\frac{\displaystyle\int_{0}^{\infty}dk\frac{\cos(kx_0)}{
s+K_{\alpha}\left\vert k \right\vert+K_{B}k^2}}{\displaystyle\int_{0}^{\infty}dk
\frac{1}{s+K_{\alpha}\left\vert k \right\vert+K_{B}k^2}}=\frac{I_2}{I_1},
\end{equation}
where $I_1$ and $I_2$ can be computed or taken from Prudnikov \cite{prudnikov}
\begin{eqnarray}
I_1&=&\frac{1}{K_B(k_2-k_1)}\ln\left\vert\frac{k_2}{k_1}\right\vert,\\
\nonumber
I_2&=&\frac{1}{K_B(k_2-k_1)}\left[\mathrm{ci}(x_0k_2)\cos(x_0k_2)-\mathrm{ci}(
x_0k_1)\cos(x_0k_1)\right.\\
&&+\left.\mathrm{si}(x_0k_2)\sin(x_0 k_2)-\mathrm{si}(x_0k_1)\sin(x_0k_1)\right],
\end{eqnarray}
where $k_{1,2}=\frac{K_1\mp\sqrt{K_1^2-4K_B s}}{2K_B}$, $\mathrm{ci}(z)$ and
$\mathrm{si}(z)$ are integral cosine and sine, respectively, defined via
$\mathrm{si}(z)=-\int_z^{\infty}\frac{\sin y}{y}dy$ and $\mathrm{ci}(z)=-\int_z
^{\infty}\frac{\cos y}{y}dy$. Thus, for $\alpha=1$, $v=0$ and
\begin{eqnarray}
\nonumber
\wp_{\mathrm{fa}}(s)&=&\frac{1}{\ln\left|\frac{k_2}{k_1}\right|}\Big[ci(x_0k_2)
\cos(x_0k_2)-ci(x_0k_1)\cos(x_0k_1)\\
\nonumber
&&+\mathrm{Si}(x_0k_2)\sin(x_0 k_2)-\mathrm{Si}(x_0k_1)\sin(x_0k_1)\\
\nonumber
&&+\pi\cos\left(\frac{k_1+k_2}{2}\right)\sin\left(\frac{k_2-k_1}{2}\right)\Big]\\
&=&\frac{1}{\ln\left|\frac{k_2}{k_1}\right|}\Big[\mathrm{ci}(x_0k_2)\cos(x_0k_2)
-\mathrm{ci}(x_0k_1)\cos(x_0k_1)\\
\nonumber
&&+\mathrm{si}(x_0k_2)\sin(x_0 k_2)-\mathrm{si}(x_0k_1)\sin(x_0k_1)\Big].
\end{eqnarray}
In the first expression an alternative way to represent results is used (which
corresponds to Mathematica) with
\begin{equation}
\mathrm{Si}(z)=\int_0^{z}\frac{\sin y}{y}dy=\frac{\pi}{2}+si(z).
\end{equation}

\section{Asymptotical behaviour of the search efficiency for $p\to\infty$}
\label{asymptotic}

In the limit $p\to\infty$ the search efficiency can be expressed through the
Brownian efficiency $\mathcal{E}_B$ (i.e., for $p=0$) in the form
\begin{eqnarray}
\mathcal{E}\left(p\right)=\frac{\mathcal{E}_B}{2}\int_0^\infty \wp_{\mathrm{fa}}
(p,s)d(st_{B}),
\label{Ep}
\end{eqnarray}
where $\wp_{\mathrm{fa}}(p,s)$  is determined by Eq. (\ref{pfp}).

\subsection{$\alpha>1$}

For $\alpha>1$ LFs have a finite search reliability. In the limit $p\to\infty$
the LFs dominate the search process, and we thus necessarily recover expression
(\ref{eff_fin}).

\subsection{$\alpha<1$.}

In the case $\alpha<1$ convergence at infinity is due to the term $k^2$
and we cannot neglect it so easily as we did for $\alpha>1$. Let us change
the variables in (\ref{Ep}) as $st_B=p^{\nu}u,\,k=p^\mu \kappa$, where $\nu$
and $\mu$ will be specified below. Then from (\ref{Ep}) we get 
\begin{equation}
2\frac{\mathcal{E}(p)}{\mathcal{E}_B}=p^\nu\int_0^\infty du
\frac{\displaystyle\int_0^{\infty}\frac{\cos (p^{\mu}\kappa)}{p^\nu
u+p^{1+\alpha\mu}\kappa^\alpha+p^{2\mu}\kappa^2}p^\mu
d\kappa}{\displaystyle\int_0^{\infty}\frac{1}{p^\nu
u+p^{1+\alpha\mu}\kappa^\alpha+p^{2\mu}\kappa^2}p^\mu d\kappa}.
\label{split2}
\end{equation}
We choose $\nu$ and $\mu$ such that
\begin{equation}
\nu=1+\alpha\mu=2\mu,
\end{equation}
i.e.
\begin{equation}
\mu=\frac{1}{2-\alpha}, \qquad\nu=\frac{2}{2-\alpha}.
\end{equation}
Then Eq.~(\ref{split2}) takes the form
\begin{eqnarray}
2\frac{\mathcal{E}(p)}{\mathcal{E}_B}=p^\nu\int_0^\infty du \frac{\displaystyle
\int_0^{\infty}\frac{\cos (p^{\mu}\kappa)}{u+\kappa^\alpha+\kappa^2} d\kappa}{
\displaystyle\int_0^{\infty}\frac{1}{ u+\kappa^\alpha+\kappa^2}d\kappa}.
\label{split2a}
\end{eqnarray}
The integral in the denominator converges at all positive $u$, does not depend
on $p$ and has an upper bound at $u=0$,
\begin{eqnarray}
\nonumber
f(u)&=&\int_0^{\infty}\frac{1}{ u+\kappa^\alpha+\kappa^2}d\kappa\le f(0)\\
&=&\int_0^{\infty}\frac{1}{ \kappa^\alpha+\kappa^2}d\kappa.
\label{bound}
\end{eqnarray}
As for the integral in the numerator, since $p\gg1$ the main contribution
comes from small $\kappa$. We thus neglect $\kappa^2$ in comparison with
$\kappa^\alpha$ and use the approach from \ref{PureLevyEff}. Hence, for the
efficiency we get
\begin{eqnarray}
\nonumber
2\frac{\mathcal{E}(p)}{\mathcal{E}_B}&\approx& p^\nu\int_0^\infty \frac{du}{f(u)}
\int_0^{\infty}\frac{\cos (p^{\mu}\kappa)}{u+\kappa^\alpha} d\kappa\\
&\sim& p^\nu \int_0^\infty \frac{du}{f(u)}\int_0^\infty d\tau e^{-u\tau}\frac{1}{
\tau^{1/\alpha}}l_\alpha\left(\frac{p^\mu}{\tau^{1/\alpha}}\right),
\end{eqnarray}
where
\begin{eqnarray}
\nonumber
y=p^{\mu}/\tau^{1/\alpha},\qquad\tau=p^{\mu\alpha}/y^{\alpha},\qquad d\tau=-
\frac{1}{\alpha}\frac{p^{\mu\alpha}}{y^{\alpha+1}}dy.
\end{eqnarray}
Thus 
\begin{eqnarray}
\nonumber
2\frac{\mathcal{E}(p)}{\mathcal{E}_B}&\sim& p^{\nu-\mu+\mu\alpha}\int_0^\infty
\frac{du}{f(u)}\int_0^\infty dy \exp\left(-\frac{up^{\mu\alpha}}{y^{\mu\alpha}}
\right)y^{-\alpha}l_\alpha\left(y\right)\\
\nonumber
&=&p^{\nu-\mu+\mu\alpha} \int_0^\infty dyl_\alpha\left(y\right) y^{-\alpha}\int
_0^\infty\frac{du}{f(u)}\exp\left(-\frac{up^{\mu\alpha}}{y^{\mu\alpha}}\right)\\
\nonumber
&=&p^{\nu-\mu} \int_0^\infty dyl_\alpha\left(y\right) \int_0^\infty \frac{e^{-t}
dt}{\displaystyle f\left(\frac{y^\alpha t}{p^{\mu\alpha}}\right)}\\
&\sim& p^{\nu-\mu}\sim p^{1/(2-\alpha)},\qquad @ p\rightarrow\infty,
\end{eqnarray}
due to relation (\ref{bound}), i.e., for $\alpha<1$
\begin{equation}
\mathcal{E}(p)\sim p^{\frac{1}{2-\alpha}}.
\end{equation}

\subsection{$\alpha=1$}

The case $\alpha=1$ requires a special treatment. The efficiency in this case is
\begin{eqnarray}
\frac{2\mathcal{E}(p)}{\mathcal{E}_B}=\int_0^\infty d(st_{B})
\frac{\displaystyle\int_0^{\infty}\frac{\cos
kdk}{st_B+pk+k^2}}{\displaystyle\int_0^{\infty}\frac{dk}{st_{B}+pk+k^2}}.
\end{eqnarray}
Making the same change of variables as in Appendix F.2, we get (cf. (F.5))
\begin{eqnarray}
\frac{2\mathcal{E}(p)}{\mathcal{E}_B}=p^2\int_0^\infty du
\frac{\displaystyle\int_0^{\infty}\frac{\cos(p\kappa)d\kappa}{u+\kappa+\kappa^2}}{
\displaystyle\int_0^{\infty}\frac{d\kappa}{u+\kappa+\kappa^2}}.
\label{asymptalphaone}
\end{eqnarray}
To proceed we start with the evaluation of the integral in the denominator
\begin{eqnarray}
\nonumber
f(u)&=&\int_0^{\infty}\frac{1}{u+\kappa+\kappa^2}d\kappa=\int_0^{\infty}\frac{1}{
\left(\kappa+\frac{1}{2}\right)^2+u-\frac{1}{4}}d\kappa\\
&=&\left\{\begin{array}{ll}f_1(u)=\frac{1}{2\sqrt{1/4-u}}\ln\left\vert\frac{1/2+
\sqrt{1/4-u}}{1/2-\sqrt{1/4-u}}\right\vert,& u<1/4,\\
f_2(u)=\frac{1}{\sqrt{u-1/4}}\left(\pi/2-\arctan\frac{1}{2\sqrt{u-1/4}}\right
),& u>1/4,\\
2,& u=1/4,\end{array}\right..
\end{eqnarray}

For the integral in the numerator similar to the case $\alpha<1$ (Appendix F.2)
the main contribution comes from small $\kappa$ values due to $p\gg1$. Hence we
can neglect $\kappa^2$ in comparison with $\kappa$. Thus
\begin{eqnarray}
f(u)=\int_0^{\infty}\frac{\cos(p\kappa)}{u+\kappa+\kappa^2}d\kappa\underset{p\gg1}
\simeq\int_0^{\infty}\frac{\cos(p\kappa)}{u+\kappa}d\kappa=g(pu),
\end{eqnarray}
where g(z) can be expressed through sine and cosine integrals $Si(z)$ and
$Ci(z)$
\begin{equation}
g(z)=-Ci(z)\cos(z)-\left(Si(z)-\pi/2\right)\sin(z).
\end{equation}
Eq.~(\ref{asymptalphaone}) can be rewritten as
\begin{equation}
\frac{2\mathcal{E}(p)}{\mathcal{E}_B}=p^2\int_0^{1/4}
\frac{du}{f_1(u)}g(pu)+p^2\int_{1/4}^{\infty}\frac{du}{f_2(u)}g(pu).
\label{splitalphaone}
\end{equation}
For the second term in the latter expression one can use an asymptotic of $g(z)
\sim1/z^2$ for $pu\gg1$ since $p\gg1$ (see Eq. 5.2.35 in \cite{Abramowitz}). This
implies that the contribution from the second term does not grow with increasing
$p$ at large $p$ values.

The first term can be rewritten as
\begin{equation}
p^2\int_0^{1/4} \frac{du}{f_1(u)}g(pu)=p\int_0^{p/4} \frac{dy}{f_1(y/p)}g(y).
\end{equation}
The upper bound of this term is given by  
\begin{equation}
p\int_0^{p/4} \frac{dy}{f_1(y/p)}g(y)<\frac{p}{f_1(1/4)}\int_{0}^{\infty}dyg(y),
\end{equation}
as $g(y)$ is integrable on $[0,\infty)$ and we can replace the upper limit
$p/4$ of the integral with $\infty$ at $p\gg1$. Thus, the first term in
Eq.~(\ref{splitalphaone}) does not grow faster than $p$. To get a lower bound for
the growth limit of large $p$ we use the first mean value theorem \cite{Ryzhik}
and a small argument asymptotic $f_1(u)\sim-\ln u$, yielding
\begin{equation}
p\int_0^{p/4}\frac{dy}{f_1(y/p)}g(y)=\frac{p}{f_1\left(y^*/p\right)}\int_0^{p/4}
dyg(y)\sim\frac{p}{-\ln\left(y^*/p\right)}\int_0^\infty dy g(y),
\end{equation}
where $0<y^*<p/4$. Hence
\begin{equation}
\frac{\mathcal{E}(p)}{\mathcal{E}_B}\sim\frac{p}{\ln p},
\end{equation}
which is confirmed by numerical simulations.

\ack

VVP wishes to acknowledge very fruitful discussions with I. Sokolov. RM
acknowledges funding from the Academy of Finland within the Finland
Distinguished Professor programme.

\end{document}